\documentclass[fleqn,twoside]{article}
\catcode`\@=11

\def\noi{\noindent}
\usepackage[dvips]{graphicx}

\renewcommand{\section}{\@startsection{section}{1}{0pt}%
        {-3.5ex plus -1ex minus -.2ex}{2.3ex plus .2ex}%
        {\large\bf\protect\raggedright}}

\renewcommand{\subsection}{\@startsection{subsection}{2}{0pt}%
        {-3ex plus -1ex minus -.2ex}{1.4ex plus .2ex}%
        {\normalsize\bf\protect\raggedright}}

\renewcommand{\thesubsubsection}%
        {\arabic{section}.\arabic{subsection}.\arabic{subsubsection}.}
\renewcommand{\@oddhead}%
{\raisebox{0pt}[\headheight][0pt]{%
    \vbox{\hbox to\textwidth{\rightmark \hfil \rm \thepage \strut}\hrule}}}
\renewcommand{\@evenhead}%
{\raisebox{0pt}[\headheight][0pt]{%
   \vbox{\hbox to\textwidth{\thepage \hfil \leftmark \strut}\hrule}}}
\newcommand{\heads}[2]{\markboth{\protect\small\it #1}{\protect\small\it #2}}
\newcommand{\Acknow}[1]{\subsection*{Acknowledgement} #1}

\newcommand{\Arthead}[3]{ \setcounter{page}{#2}\thispagestyle{empty}\noi
    \unitlength=1pt \begin{picture}(500,40)
        \put(0,58){\shortstack[l]{\small\it Gravitation \& Cosmology,
                        \small\rm Vol. 5 (1999), No. #1, pp. #2--#3\\
        \footnotesize \copyright \ 1999 \ Russian Gravitational Society} }
    \end{picture}          }
\newcommand{\Title}[1]{\noi {\Large #1} \\}

\newcommand{\Authors}[4]{\noi
        {\large\bf #1\dag\ #2\ddag}\medskip\begin{description}
        \item[\dag]{\it #3} \item[\ddag]{\it #4}\end{description}}
\newcommand{\Rec}[1]{\noi {\it Received #1} \\}

\newcommand{\Abstract}[1]{\vskip 2mm \begin{center}
        \parbox{16.4cm}{\small\noi #1} \end{center}\medskip}

\newcommand{\Ref}[1]{Ref.\,\cite{#1}}

\def\nqq{\hspace*{-2em}}
\def\nhq{\hspace*{-0.5em}}

\def\cm{\hspace*{1cm}}
\def\inch{\hspace*{1in}}

\def\al{&\nhq}
\def\lal{&&\nqq {}}
\def\eq{Eq.\,}

\def\beq{\begin{equation}}
\def\eeq{\end{equation}}
\def\bear{\begin{eqnarray}}
\def\bearr{\begin{eqnarray} \lal}
\def\ear{\end{eqnarray}}
\def\earn{\nonumber \end{eqnarray}}

\def\nnn{\nonumber\\ \lal }

\def\eql{\al =\al}

\def\tst{\textstyle}

\def\fract#1#2{{\tst\frac{#1}{#2}}}

\def\half{{\fract{1}{2}}}

\def\d{\partial}

\def\im{\mathop{\rm Im}\nolimits}

\def\In{{\rm in}}
\def\out{{\rm out}}
\def\mpl{m_{\rm pl}}
\begin{document}
\heads{V.A. Berezin, A.M. Boyarsky and A.Yu. Neronov}
      {On the Mechanism of Hawking Radiation}

\twocolumn[
\Arthead{00}{1}{10}

\vspace*{-5mm}

\Title {ON THE MECHANISM OF HAWKING RADIATION}

   \Authors{V.A. Berezin, A.M. Boyarsky}
           {and A.Yu. Neronov}
           {Institute for Nuclear Research
            of the Russian Academy of Sciences\\
            60-th October Anniversary Prosp., 7a,
            117312, Moscow, Russia}
           {Theoretische Physik, Universit\"at M\"unchen,
            D-80333, M\"unchen, Germany}

\Rec{20 March 1999}

\Abstract
{In our previous papers the first step was made to the
construction of a global wave function on the configuration space of a
self-gravitating shell.  The asymptotic behaviour of analytical
wave functions at the infinities was analyzed. As a result, a discrete mass
spectrum of a quantum black hole and a discrete spectrum for the Hawking
radiation were found.  In the present paper we study a global quasiclassical
solution inside and outside the horizon. The result is rather unexpected:
for a quasiclassical solution with two waves of equal amplitudes under
the horizon we obtain, in the outer region of the black hole, ingoing
and outgoing waves with the amplitudes $Z_{\In}$ and $Z_{\out}$ such that
$Z_{\In}^2/Z_{\out}^2 = \exp\{-{\delta A}/(4\mpl^2)\}$
where $A$ is the black hole horizon area.
This result exactly coincides with the main result of the Hartle and
Hawking consideration \cite{h-h}, from which one can derive the value of
the black hole temperature and entropy.}
] 

\section{Introduction}

Conventional quantum field theory is a set of quantum mechanics
for $N$-particle states and transitions between them.
Its starting point is a Fock space
spanned by the states of (approximately) free particles with different values
of energies and momenta. Having at first a quantum mechanical description
of these particles, we construct their Hilbert space and then take into
account the processes of creation and annihilation, thus secondary
quantizing the theory. But it is now well known that $N$-particle states at
the Planckian scale could be rather different from free particle states.
Indeed, there is no free particle state for particles with trans-planckian
energies moving in different directions. The Compton wavelength in this case
is smaller than the gravitational radius of the particle.  Such particles
will inevitably form black holes (see e.g. \cite{thooft}).

So it is natural, when dealing with quantum field theory at the Planckian
scale, first to try to construct a Hilbert space containing $N$ gravitating
particles before second quantization.

At first sight, this $N$-particle quantum mechanics is introduced only as a
technical tool regularizing quantum field theory at the Planckian scale.
This could be done in different ways. In particular, there could be
different models of quantum mechanics of gravitating particles.  But
fortunately there is a language for studying the physical content of these
models at already the quantum-mechanical level.  Namely, since the ``black
hole states'' should be present in Hilbert space, this first-quantized
quantum mechanics should be suitable for describing such physical phenomena
as the Hawking radiation, mass spectrum, black hole entropy.  Analyzing
certain predictions about these phenomena (like the Hawking radiation
spectrum or the black hole entropy formula) given by different quantum
mechanics, one could choose among them. But in gravitational models we
have some special property: the gravitational field is not just the field
responsible for the gravitational interaction between particles, but it also
determines the global structure of the space-time. Usually the space-time
itself is the configuration space for particle dynamics. So, in general,
when we take into account the gravitational interaction, the structure of
the ``configuration space'' becomes dependent on the initial conditions, as
could be observed by analyzing classical solutions. Hence it is impossible
to define a configuration space for gravitating particles naively. On the
quantum-mechanical level another problem arises.  In quantum mechanics
we are supposed to have a superposition of different classical states. It
means that in the case of a self-gravitating system we are forced to work in
terms of superposition of different space-time geometries. There have been
plenty of attempts to describe this picture but it still remains a puzzle
how to construct any field theory on such ``quantum space-time''.

This problem is the most important if we want to construct a field
theory taking into account the gravitational interaction at the Planckian
scale. As a matter of fact, it has very little to do with the problem of
scattering, creation and annihilation of gravitons. Being present in field
theory, the problem of ``quantum space-time'' arises already at the
quantum-mechanical level, i.e. for systems with a finite number of degrees
of freedom.  Thus we can try to solve it in this simplest case when it is
easier to obtain a correct definition of the configuration space.

How can we construct a quantum mechanics of gravitating particles without
facing the problems of creation and annihilation of gravitons?  A possible
approach to this problem is to take into account only some global
degrees of freedom of the gravitational field relevant for the dynamics of
gravitating particles. These degrees of freedom are topological in the
absence of sources.

It is well known that gravitational theories  contain
topological degrees of freedom. For example, 2+1 gravity is pure topological
and equivalent to the Chern-Simons theory with the gauge group
$ISO(2,1)$ \cite{2+1}. Spherically symmetric 3+1 gravity without matter is
also topological due to the Birkhoff uniqueness theorem.

On the other hand, in some complicated cases,
when the theory is not topological,
one can look for its topological sectors.
If we are interested only in some particular observables, it could
be enough to study quantum mechanics of these global degrees of freedom.
For example, we  could separate the dynamics of global degrees of freedom
from the dynamics in volume by extracting some
important surface terms from the action functional
thus obtaining a field theory on the boundary surface\footnote%
        {This idea is in good agreement with the so-called 't Hooft
        ``holographic principle'' which states that there should exist some
        surface, 2+1-dimensional, field theory, whose degrees of freedom are
        relevant to the black hole entropy (\cite{holpr}).  An attempt to
        describe these quantum black hole degrees of freedom by a
        Chern-Simons field theory on the horizon (or at infinity) was
        recently made in \cite{ashtekar, carlip, solodukhin}.  For a recent
        review see \cite{banados}.}.

The corresponding field theories in such topological models should have a
finite number of physical degrees of freedom. Thus the quantum-mechanical
phase space is present in the problem from the very beginning. Then one has
to add some sources and the proper Hamiltonian which results in the
transition from topological to physical theory, as could be easily
seen, for example, in the 2D Yang-Mills case (cf. \cite{gorsk-nekr} and
refs. therein).  As a result, one has a quantum-mechanical problem (it is
often integrable) describing the dynamics of these topological degrees of
freedom.  Mathematically this finite-dimensional system is obtained by
Hamiltonian reduction.

Let us try to apply this approach to gravitating particles.
To construct a quantum mechanics for such a topological field theory with a
particle we need first to clarify what is the ``classical phase space".
The latter can be defined as the space of all classical solutions modulo
gauge transformations. For the case of self-gravitating particles in $2+1$
dimensions this approach was used in \cite{matchull-welling}.

In what follows (as in our previous papers)
we will study maybe the simplest gravitational model of the type described
above --- spherically-symmetric gravity with a self-gravitating thin
dust shell (see \cite{ber-kuz, prd57, prl}).

\begin{figure}
  \includegraphics[width=\columnwidth]{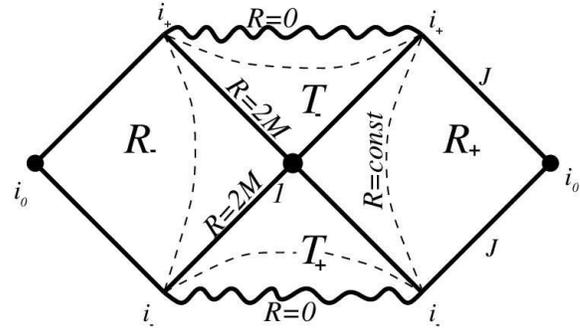}%
  \caption{Penrose diagram for a Schwarzchild black hole.
    Dashed lines are curves of constant radius}
    \label{fig:kruskal}
\end{figure}

\begin{figure}
  \includegraphics[width=\columnwidth]{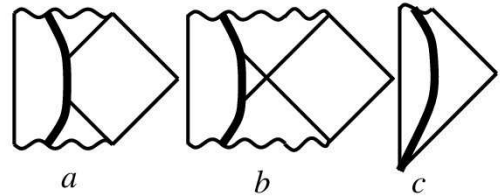}
    \caption{Different space-times with a self-gravitating shell}
    \label{fig:trajectories}
\end{figure}

In this model we see among the classical solutions space-time manifolds with
different geometries.  In Fig.\,\ref{fig:kruskal} the geometry of the
complete Schwarzschild space-time is shown. It contains two isometric
regions with two singularities (future and past) at $R=0$ and two infinities
in the left and right asymptotically flat regions $R_+$ and $R_-$.  In Fig.
\ref{fig:trajectories} different types of the corresponding Carter-Penrose
diagrams are presented.  In the ``black hole case" (a) the turning point of
the shell lies in the $R_+$-region and at this point it can be seen by an
observer at the  right infinity. In the ``wormhole case" (b) the turning
point of the shell is on the opposite side of the Einstein-Rosen bridge in
the $R_-$-region and could not be seen from $R_+$-infinity.  In the case of
unbounded motion (c) the shell starts from the infinity $R=\infty $ in the
$R_+$-region and collapses to $R=0$ forming a singularity.

All these classical configurations should be present in the
finite-dimensional phase space of the gravitating shell.

Due to the high symmetry it is possible in this model to fulfil the  above
reduction explicitly. In the case of spherically symmetric gravity without
matter such a reduction was made by K. Kucha\v{r} (\cite{kuchar}). The
resulting ``topological'' gauge-invariant degree of freedom is a variable
$m$ (and the corresponding momentum) which is defined by the boundary
conditions at the infinity and is nothing but the Schwarzschild mass
measured by an observer at infinity. If a thin dust shell is included
as a source, we have (after reduction) another ``physical'' degree of
freedom, describing the shell motion. The variable $m$ enters into the
resulting equation for the gravitating shell and can be formally considered
as a parameter. There is another parameter, the bare mass of the shell $M$.
In classical mechanics, cases (a) and (b) are realized for bounded
motion of the shell when $M/m< 2$ and $M/m> 2$, respectively.

There were several attempts to construct both classical and quantum mechanics
for such a system. For example, in \cite{prd56} the reduced phase space for
a self-gravitating shell was constructed as a set of initial data for
the black hole case (a) only. In \Ref {kr-wilch} the local wave function of
a self-gravitating null shell was found to describe the effects of
back reaction in non-thermal corrections to the spectrum of the Hawking
radiation. But in quantum mechanics the wave function should be defined over
the whole configuration space and it is of crucial importance to construct
a global picture for the dynamics of the system in hand. Some important
results, such as the quantization conditions, can be obtained only from
global properties of the wave function. So it is necessary in our case to
construct a global configuration space taking into account all classical
solutions, in particular, the ``wormhole'' case (b) as well as the ``black
hole'' case (a) for any value of the ratio $M/m$. But for each particular
value of this ratio the part of the global configuration space representing
classical solutions with the opposite sign of $(M/m-2)$ should be
classically forbidden in an effective quantum mechanics of the
self-gravitating shell. For example, in the black hole case $(M/m<2)$ the
part of the configuration space representing wormhole classical solutions is
classically forbidden.

This is the first qualitative difference of the self-gravitating shells
motion from test particle (shell) motion on a fixed Kruskal background when
the shell can move in all parts of the Carter-Penrose diagram irrespective
of the value of $M/m$. The appearance of an additional classically forbidden
region in the configuration space, where the wave function should
exponentially decrease, results in a new quantization condition for the
parameters $M$ and $m$ (\cite{prd57,prl}). This effect (whose physical
consequences are discussed in \cite{prl} and will be discussed below) can be
illustrated by the following simple quantum-mechanical example. Let us
consider the following radial Schr\"odinger equation:
\bearr
\frac{d^2R(r)}{dr^2}+\frac{2}{r}\frac{dR(r)}{dr}+\frac{2m}{\hbar^2}\left( E-
        \frac{A}{r^2}+\frac{B}{r}\right) R(r)=0. \nnn
\ear
Then, let us suppose that the asymptotic behaviour of the wave function
at negative infinity $r\to -\infty$ along the real line is also
important for some physical reasons (for example, the true configurations
variable is the area $s=r^2$, and the classical configuration space is a
positive semi-axis $s>0$). In this case it is easy to see from the exact
solution (see \cite{prl}) that this new requirement, together with usual ones
at $r\to \infty$ and $r\to 0$, gives not only a quantization
condition for the parameter $E$, but also one more quantization condition,
so that the parameter $A$ is quantized as well.

\begin{figure}
  \includegraphics[width=\columnwidth]{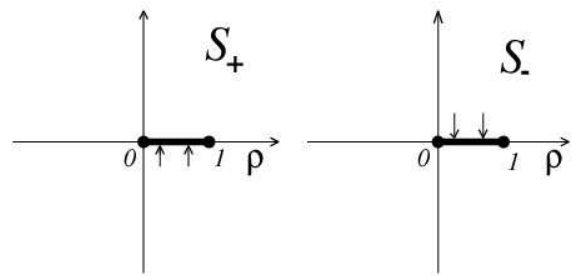} %
    \caption{The real section of the Riemannian surface
      $S_F$ (\ref{eq:sf}) covers the real line twice}
 \label{fig:sf}
\end{figure}

In \cite{prd57} the formalism was constructed to describe global properties
of the configuration space and globally-defined quantum mechanics for
the case of a self-gravitating shell. In this case, after reduction, using
the Kucha\v{r} gauge-invariant variables, we are left with the only
nontrivial equation describing the dynamics of the shell. Formally this
dynamics is one-dimensional. The variable $\hat R$ which describes the
position of the shell is gauge-invariant and has the meaning of the shell
radius.  But to parameterize the whole configuration space it is not enough
to have $R$ varying from $0$ to $\infty$. This  can be easily seen from the
observation that it is impossible to distinguish black-hole-type
classical solutions from the wormhole-type solutions in terms of
the variable $R$ (see Fig.\,\ref{fig:trajectories}).
This variable covers the configuration space twice.
Fortunately there is a way to avoid this difficulty.
The equation which governs the dynamics of the shell is an
equation in finite differences.
The shift of the argument of the wave function in this equation occurs
along the imaginary axis, and this means that the equation is actually
defined not on the real line, but over some complex manifold. The equation
for the shell dynamics contains the square root of the  ``Schwarzschild
factors'' $\sqrt{1-\frac{2m}{R}}$.  So the natural complex manifold for the
equation is the Riemannian surface $S_F$ on which the coefficients of the
equation are analytical functions. This Riemannian surface is just a
two-(real)dimensional sphere obtained after gluing two complex planes along
the sides of the cuts made on each plane along the interval between
the branching points of the coefficients on the real line.  The
configuration space for our self-gravitating shell is the real section of
$S_F$. This configuration space properly represents different classical
solutions for the self-gravitating shell (for a detailed analysis see
\cite{prl} and \cite{grr}).  All main results from our models are due to the
non-trivial structure of the configuration space.  This real section
$\im(\rho)=0$ of the Riemannian surface $S_F$ covers the real line twice,
as is shown in Fig.\,\ref{fig:sf}. It consists of the $V_{\pm}, T_{\pm}$ and
$R_{\pm}$ intervals. The sign of the branching function is taken to be $+$ on
the intervals $V_+$, $R_+$ and $-$ on the intervals $T_-$, $R_-$ and
$(F_{\In}F_{\out})^{1/2}=\pm i\sqrt{\left|F_{\In}F_{\out}\right|}$ on
$T_\pm$, respectively.

In \cite{prd57,prl,exact} the first step to the construction of a global
wave function on such a complicated configuration space was made.
The asymptotic behaviour of the analytical wave functions at the  infinities
in $R_{+}$ and $R_{-}$ regions was analyzed. As a result, a discrete mass
spectrum of bound states and a discrete spectrum for infinite motion of the
system were found. Analyzing these two spectra, the Bekenstein-Mukhanov mass
spectrum for black holes \cite{bekenstein} was obtained.

In the present paper we study a global quasiclassical solution for
the $V_{+}$, $T_{+}$ and $R_{+}$ regions. The result is rather unexpected
--- for a quasiclassical solution with two waves of equal amplitudes under
the horizon we obtain, after analytical continuation in the $R_{+}$ region,
ingoing and outgoing waves with the amplitudes $Z_{\In}$ and
$Z_{\out}$.  Namely,
\[
        Z_{\In}^2/Z_{\out}^2=\exp\{-\delta A/(4\mpl^2)\}
\]
where $A$  is the black hole horizon area. This exactly coincides with the
main result of the Hartle and Hawking consideration \cite{h-h} from which
one can derive the values of the black hole temperature and entropy.

\section{Quantum mechanics of self-gravitating massless particles}

As was shown in \cite{prd57,prl}, the radial relativistic Schr\"o\-din\-ger
equation for the massless self-gravitating null-dust shell has the form
\beq
        \label{eq:main}
        \Psi(S+i\zeta)+\Psi(S-i\zeta)=\frac{F_{\In}+F_{\out}}
                                        {\sqrt{F_{\In}F_{\out}}}\Psi.
\eeq
Here
\beq
        S= R^2\Big/ (4G^2m^2)
\eeq
is a dimensionless variable which
measures the area of the shell ($G$ is the gravitational constant and
$m=m_{\out}$ is the Schwarzschild mass of the black hole as seen by
an observer at infinity). The dimensionless shift parameter is
\beq
\zeta=\mpl^2/(2m^2);\qquad \mpl =\sqrt{\hbar c/G}
\eeq
($\mpl $ is the Planck mass). The functions $F_{\In,\out}$ are just the
coefficients of the Schwarzschild metric inside and outside the shell:
\bear
     F_{\out}=1- 1\Big/ \sqrt{S}, \qquad  F_{\In}=1-\mu\Big/ \sqrt{S}
\ear
($\mu=m_{\In}/m_{\out}$ is the quotient of the Schwarzschild masses inside
and outside the shell). If we suppose that the energy of the null shell
\beq                           \label{eq:eps}
        \epsilon =m_{\out}-m_{\In}
\eeq
is much smaller than the black hole mass $m_{\out}$, then
\beq
m_{\In}\approx m_{\out}=m; \quad\mbox{ and }\quad \mu\approx 1-\epsilon/m .
\eeq
So in the limit of test particles (when the back reaction is not taken into
account) we can expand all the quantities in powers of the small parameter
$\epsilon / m$.

Another limiting situation is when the shift parameter is small compared to
a characteristic scale on which the wave function varies significantly, then
one can approximate the shifted wave function $\Psi(S\pm i\zeta)$ by its
Tailor expansion near $S$, so that \eq (\ref{eq:main}) takes the form
\beq
  \label{eq:ode}              -\zeta^2\Psi''(S)+2\Psi(S)
        =\frac{F_{\In}+F_{\out}}{\sqrt{F_{\In}F_{\out}}}\Psi(S),
\eeq
which is just the usual Schroedinger equation of nonrelativistic quantum
mechanics.  This is natural because the limit $\zeta\to 0$ (or $\mpl \to
0$) is either nonrelativistic, or classical, or the limit of weak
gravitational field. Thus we see that the coefficient in the right-hand side
of \eq (\ref{eq:main}) plays the role of a potential term for the motion
of the null shell in a gravitational field.

One important note is that the ``truncated'' equation (\ref{eq:ode}) is
certainly not valid in the vicinity of the in- and out- horizons --- they
are singular points of \eq (\ref{eq:main}) and we cannot assume that
$\Psi$ varies slowly there.

The shift of the argument of the wave function in \eq (\ref{eq:main}) is
along the imaginary axis, this means that the equation is actually defined
not on the real line, but over some complex manifold.  The natural complex
manifold for \eq (\ref{eq:main}) is the Riemannian surface $S_F$ of
the branching function
\beq                                                 \label{eq:sf}
  (F_{\In}F_{\out})^{1/2}
                =\frac{\left\{(\rho-1)(\rho-\mu)\right\}^{1/2}}{\rho}
\eeq
($\rho=\sqrt{S}$)
on which the coefficients of the equation are analytical functions.
This Riemannian surface $S_F$ is just a two-(real)dimensional sphere
obtained after gluing two complex planes along the sides of the cuts
made on each of them along the interval $\rho\in (\mu ,1)$ of the real line.
The configuration space for our self-gravitating shell
is the real section of  $S_F$. This configuration space represents
properly a different classical solution for a self-gravitating shell
(for a detailed analysis see \cite{prl} and \cite{grr}).
All main results in our models are due to a non-trivial structure
of the configuration space. This real section $\im(\rho)=0$ of the
Riemannian surface $S_F$ covers the real line twice, as is shown in Fig.
\ref{fig:sf}. It consists of the intervals $V_{\pm}, T_{\pm}$ and $R_{\pm}$.
The sign of the branching function (\ref{eq:sf}) is taken to be $+$ on
the intervals $V_+,R_+$, $-$ on intervals $T_-,R_-$, and
$(F_{\In}F_{\out})^{1/2}=\pm i\sqrt{\left|F_{\In}F_{\out}\right|}$ on
$T_\pm$, respectively.

\section{Quasiclassical wave function}

In our previous papers the main attention was devoted
to the behaviour of the wave function at the infinities in the $R_+$ and $R_-$
regions. Here we study the properties of the quasiclassical
wave function near the horizons $s=1$ and $s=\mu$.
Consider the quasiclassical solutions of \eq (\ref{eq:main}) in the form
\beq
\label{eq:anzatz}
\Psi =\exp\left\{\frac{i\Omega (S)}{\zeta}\right\}(\phi_0+\zeta\phi_1+
                \ldots).
\eeq
Substituting (\ref{eq:anzatz}) into (\ref{eq:main}), one gets in the zero
order in $\zeta$ the Hamilton-Jacobi equation for $P_S=\d \Omega/ \d S$:
\beq
\label{eq:hj}
\cosh\{P_S\}=\frac{F_{\In}+F_{\out}}{2\sqrt{F_{\In}F_{\out}}}
\eeq
whence it follows
\beq                               \label{eq:ps}
P_S =\pm\ln\left(\frac{F_{\out}}{F_{\In}}\right)^{1/2}=\pm\ln\left(
\frac{\sqrt{S}-1}{\sqrt{S}-\mu}\right)^{1/2}.
\eeq
The $\pm$ signs correspond to expanding and collapsing trajectories of the
null shell.

The points $S=S_0$ where $P_S=0$ are the turning points of the shell
classical motion. From (\ref{eq:ps}) it is easy to see that there
are no such points for the null-shell motion. But the points $S=1$ and
$S=\mu^2$ (the points where the apparent horizons of the internal and
external Schwarzschild metrics are situated) are singular points of
\eq (\ref{eq:main}). The quasiclassical anzatz (\ref{eq:anzatz}) is
not a good approximation for solving \eq (\ref{eq:main}) near these points.

In the region $S\in (\mu^2, 1)$ between the horizons the momentum $P_S$
(\ref{eq:ps}) has an imaginary part, so this is an analogue of
the ``classically forbidden'' region if we use the analogy with
nonrelativistic quantum mechanics provided by the form (\ref{eq:ode}) of
\eq (\ref{eq:main})\footnote%
    {It is interesting to note that these singular points could be treated
    as turning points. \eq (\ref{eq:main}) is defined over the
    Riemannian surface $S_F$. The points $\mu^2$ and $1$ are branching
    points of this Riemannian surface and the coordinate $S$ is not a
    regular coordinate on $S_F$ in the neighbourhoods of these points.  The
    regular coordinate in the vicinity of $S=\mu^2$ is $u=\sqrt{\rho-\mu}$.
    The momentum conjugate to the coordinate $u$ is
\[
    P_u=P_S\,\frac{dS}{du}=\pm
        2u(u^2+\mu)\ln \frac{u}{(1-\mu)-u^2}
\]
    which is equal to zero at the point
    $u=0$. This means that in terms of the regular coordinate $u$ on
    $S_F$ the point $S=\mu^2$ is a turning point of the classical motion
    rather than a singular point of the classical dynamical system. The
    singularity of the coefficients of \eq (\ref{eq:main}) originates
    actually from the singularity of the coordinate $S$ on the Riemannian
    surface near $S=\mu^2$ which in turn is caused by an irregular
    behaviour of the radial coordinate $R=2Gm\sqrt{S}$
    on the horizon $R=2Gm_{\In}$ of the Kruskal manifold. }.
But it should be noted that this analogy is not
direct because this region is not entirely forbidden classically: of course,
we have in this region a trajectory of A particle falling to the
black hole singularity. The true origin of this ``special region''
is that it is indeed classically forbidden for a particle trajectory
going outside the black hole. In some sense, the origin of this
ban is not dynamical but casual --- a particle should go faster than light to
get out of the horizon from the black hole interior.
Quantum-mechanically this situation was analyzed in detail for the
second-order equation  (\ref{eq:ode}) (for a bare mass of the shell $M$ not
necessarily equal to zero) in \cite{prd57}. This analysis is valid in the
$T$-region far from the horizon $S=1$. The result is that we have only
an ingoing (or outgoing) wave in $T_{\pm}$-regions, respectively. The wave
in the opposite direction is enormously damped near the horizon relative to
the ``correct'' quasiclassical waves in each regions. This quasiclassical
picture reproduces the classical behaviour of the shell.

If the energy $\epsilon$ of the shell is small as compared to the mass of
black hole, then $\mu$ is close to $1$ and the region situated between
$S=\mu^2$ and $S=1$ is very narrow and the contribution of the damped waves
is not negligible.  In what follows we will try to take this contribution
into account and to determine its physical meaning.

\subsection{States inside and outside the horizon}
\label{sec:resonant}

Another important feature in the present consideration which differs from
the situation described in \cite{prd57}, when $\mu$ was equal to zero,
is the appearance of the region $V_{+}$. In this region we also have two
real solution of \eq(\ref{eq:hj}) and thus two different quasiclassical
waves. But the nature of this region is quite different from that of the
region $R_+$. We should stress that the coordinate $R$ under the horizon
(including the region $V_+$ as well!) is actually a time coordinate,
and the quasiclassical wave function
\beq                                                        \label{quasi}
        \Psi\sim \exp \biggl\{\frac{i}{\zeta}
                       \int_S  P_{\tilde S}d\tilde S\biggr\}
\eeq
with $P_S>0$ represents a wave moving forward in time, while the solution
(\ref{quasi}) with $P_S<0$ represents a wave moving backward in time.

The classical particle trajectory under the horizon which starts near $R=0$,
propagates backward in time up to the horizon, then is reflected from it and
then propagates forward in time back to the singularity $R=0$.

According to the usual interpretation of waves propagating backward in time
we might treat them as antiparticles which propagate forward in time, but
with the opposite sign of energy.  This is clear from \eq (\ref{eq:hj}).
The energy of the particle is $\epsilon=\delta m =m_{\out}-m_{\In}$.  If we
take the solution of (\ref{eq:hj}) with the minus sign before the logarithm
in the r.h.s., we can write
\beq                                                      \label{under1}
P_S=- \ln\frac{F_{\In}}{F_{\out}}=
        +\ln\frac{F_{\out}}{F_{\In}} =
        +\ln \frac{\tilde F_{\In}}{\tilde F_{\out}}
\eeq
where $\tilde F_{\In}=F_{\out}$ and $\tilde F_{\out}=F_{\In}$. This means that
$\tilde m_{\out}=m_{\In}$, $\tilde m_{\In}=m_{\out}$
and $\tilde E=-E$ --- instead of treating part of the
trajectory  as a the trajectory of a particle of energy
$E$ propagating back in time, we can treat as that of a particle of energy
$-E$ propagating forward in time. Each solution describes the situation when
either a particle or an antiparticle eventually falls into the singularity
because there is a probability flow directed to $R=0$.  Now we have to make
the first step in the construction of the global quasiclassical solution in
the configuration space: we should glue the waves in the $V_+$ region with
the waves in the $R_+$ region.  This may be done as usual by
analytical continuation through the complexified configuration space. We
will not consider below the continuation of the solutions to the $R_-$ and
$V_-$ regions, so the above continuation will actually be made through the
ordinary complex plane.

The integral in (\ref{quasi}) can be calculated explicitly
in the case in question and takes the form
\bearr
        \int_S P_{\tilde S}d\tilde S =
                        \int_x x\ln \frac {x-1} {x-\mu} dx\nnn
        = \frac{(x-a)^2}{2}\bigl[\ln(x-a)-\half\bigr] \nnn
\cm\cm
        + a(x-a)[\ln(x-a)-1] \Big|^{a=1}_{a=\mu}
\earn
(here $x=\sqrt{S}$).

We can continue this expression analytically from $V$ to the $R$ region
along the contour situated far from the branching points $x=1$
and $x=\mu$. The only result of such a continuation is the appearance
of the additional terms $i\pi$ in each logarithm:
\bearr                                                \label{delta}
        \frac{(x-a)^2}{2}(\ln(x-a)-\frac{1}{2}+ \pi i)\nnn
   \cm          + a(x-a)[\ln(x-a)-1+\pi i] \Big|^{a=1}_{a=\mu}.
\ear

As a result, we obtain that the quasiclassical wave function
(\ref{quasi}) acquires, after the analytical continuation, an additional
factor in its amplitude
\bear                                                   \label{cont}
\Psi \eql A \exp \left\{ \int_S P_{\tilde S}d\tilde S\right\}
 \nnn
\to A \exp \left\{ \int_S P_{\tilde S}d\tilde S\right\}
                \exp \biggl(\frac{\pi}{\zeta}\frac{1-\mu^2}{2}\biggr).
\ear

The second solution acquires the reversed factor
\bear                                           \label{cont1}
\Psi \eql A \exp \left\{-\int_S P_{\tilde S}d\tilde S\right\} \nnn
 \to
A\exp \left\{ -\int_S P_{\tilde S}d\tilde S\right\}
                \exp \biggl(-\frac{\pi}{\zeta}\frac{1-\mu^2}{2}\biggr).
\ear

So for a state with both waves having equal amplitudes in the $V_+$
region we have in the $R_+$ region outgoing and ingoing waves with
amplitudes related to each other as in (\ref{cont}), (\ref{cont1}).
The validity of the above consideration depends now on the following
important problem to be analyzed. The quasiclassical anzatz (\ref{quasi}) is
not a good approximation for the true wave function not only at the points
$S=1$ and $S=\mu^2$ --- the branching points of the momentum $P_S$ --- but
also at the so-called Stokes lines (see e.g. \cite{wazov}). Thse lines are
solutions of the equation
\beq                     \label{stoks}
        \im \int P_S dS=0.
\eeq
To take seriously the above analytical continuation, we must be sure that the
path in the complex plane along which we continue our quasiclassical
solution does not intersect Stokes lines. Otherwise we can lose some
important part of the quasiclassical solution which is exponentially small
compared with the wave (\ref{quasi}) before the Stokes line but becomes
large after the intersection. But fortunately in our case it can be easily
seen (analytically as well as numerically) that we can reach the region
$(1,\infty)$ from $(0,\mu^2)$ through the complex plane without
intersecting the Stokes lines.

\section{Hawking radiation spectrum}
\label{sec:hawking}

\eq (\ref{eq:main}) is a field equation for first-quantized
self-gravi\-tating massless particles in the field of a black hole.
(We suppose that it is this equation that must replace the
radial Klein-Gordon equation for the $s$-modes of a scalar
field if we want to take into account its back reaction
onto the gravitational field of the black hole.)

We suppose that the vacuum state of second quantized theory
inside the horizon consists of zero-mode oscillations of particles with
different energies. The natural property of the vacuum would be that all the
possible zero modes with different energies $\epsilon$
are present with the same amplitude. A  particle-antiparticle pair
with the energy $\epsilon$ falling into the singularity is presented
in our model as a solution which consists, under the horizon, of both
quasiclassical waves (forward and backward with respect to the variable $R$
which is time-like in the $V$-region) with equal amplitudes.

Now from the previous section we know that such a state under the horizon
gives us the ingoing and outgoing  waves with the amplitudes $Z_{\In}$
and $Z_{\out}$,
\beq                           \label{k^2}
P=\frac{Z_\In^2}{Z_{\out}^2}=\exp\biggl\{-\frac{2\pi}{\zeta}(1-\mu^2)\biggr\}.
\eeq

Let us look at the last formula in more details. We must recall the
definition of $\zeta$ and rewrite (\ref{k^2}) in a more convenient form:
\bearr                                                   \label{P}
P=
  \exp \biggl\{-\frac{2\pi^2 m_{\out}^2}{\mpl ^2}\frac{(m_{\out}^2
                                -m_{\In}^2)}{m_{\out}^2}\biggr\}   \nnn
\inch =
        \exp \biggl\{- \frac{4\pi} {\mpl^2}\frac{\delta R_g^2}{4}\biggr\}.
\ear
Introducing the area of the horizon $A$, we obtain finally
\beq                                                      \label{P1}
        P= \exp \biggl\{-\frac{1}{4}\frac {\delta A}{\mpl^2} \biggr\}.
\eeq

This result precisely coincides with the main result of Hartle and
Hawking in \cite{h-h}. Following their line of reasoning,
we can treat this probability distribution as the Gibbs distribution
\beq                                                      \label{eq:temp}
        P=\exp\{-\delta m/T\}.
\eeq
We see that it follows from the comparison of the two distributions
that the correct mass formula for the Schwarzschild  black hole is valid:
\beq
  \label{eq:mass}
  \delta m=T\,{\delta A}/4.
\eeq

Thus we have arrived at the conclusion that our first-quantized
model for a self-gravitating particle describes such an important
phenomenon of black-hole physics as the Hawking radiation.

\Acknow
{The authors are grateful to Russian Basic Research Foundation for
financial support (Grant No 99-02-18524). The work of A.B. and A.N.
was also supported by Soros Educational Programme for postgraduate students.}



\end{document}